\documentclass[10pt,a4paper]{article}
\usepackage{geometry}
\usepackage{amsmath,amssymb}
\usepackage{changepage}
\usepackage[utf8x]{inputenc}
\usepackage{textcomp,marvosym}
\usepackage{cite}
\usepackage{nameref,hyperref}
\usepackage[right]{lineno}
\usepackage{microtype}
\usepackage{graphicx}

\date{}
\title{}
\hypersetup{
 pdfauthor={Ketil Malde},
 pdftitle={},
 pdfkeywords={},
 pdfsubject={},
 pdfcreator={Emacs 25.2.2 (Org mode 9.0.9)}, 
 pdflang={English}}
\begin{document}

\vspace*{0.2in}

\begin{flushleft}
{\Large
\textbf\newline{Using sequencing coverage statistics to identify sex
  chromosomes in minke whales} 
}
\newline
\\
Ketil Malde\textsuperscript{1,2*},
Rasmus Skern\textsuperscript{1},
Kevin A.\ Glover\textsuperscript{1,3},
\\
\bigskip
\textbf{1} Institute of Marine Research, Bergen, Norway
\\
\textbf{2} Deparment of Informatics, University of Bergen, Norway
\\
\textbf{3} Department of Biology, University of Bergen, Norway
\\
\bigskip

* ketil.malde@imr.no

\end{flushleft}

\begin{abstract}
The ever-increasing number of genome sequencing and resequencing
projects is a central source of insights into the ecology and evolution
of non-model organisms.  An important aspect of
genomics is the elucidation of sex determination
systems and identifying genes on sex chromosomes.  This not only
helps reveal mechanisms behind sex determination in the
species under study, but their characteristics make sex
chromosomes a unique tool for studying the mechanisms and effects of
recombination and genomic rearrangements and how they affect adaption
and selection.  Despite this, many sequencing
projects omit such investigations.  Here, we
apply a simple method using sequencing coverage statistics to identify scaffolds belonging to
the sex chromosomes of minke whale, and show how the sex chromosome system can be
determined using coverage statistics alone.

Using publicly available data, we identify the previously unknown sex
of an Antarctic minke whale as female.  We further investigate public
sequence data from the different
species and sub-species of minke whale, and classify genomic scaffolds
from a published minke whale assembly as X or Y chromosomal
sequences.  Our findings are consistent with previous results that
identified a handful of scaffolds as sex chromosomal, but 
we are able to identify a much larger set of scaffolds, likely to represent
close to the complete sex chromosomal sequences for the minke whale.

Sequence coverage statistics provides a readily available tool for
investigating the sex determination system and locate genes on sex
chromosomes.  This analysis is straightforward and can often be performed
with existing resources.
\end{abstract}

\section{Background}
\label{sec:org61b8eb2}

Complete genome sequences are among the most important tools for
understanding the molecular inner workings and evolutionary history of
a species and its interactions with the ecosystem.
Genome sequencing projects are therefore central for broadening our
knowledge and the wealth of data they provide has become crucial to
modern biology research. 

Often, a genome assembly for a novel organism is published along with
an analysis of various aspects and characteristics of the genome.
Even for ambitious projects, resources are limited, and many
potentially interesting topics will by necessity only receive a
superficial treatment.  For instance, the recent study which published
the common minke whale \emph{(Balaenoptera acutorostrata)} genome only
performed a brief investigation of sex chromosomes based on sequence
similarity to the bovine genome \cite{yim2014minke}.  Inexpensive
analytical methods that can produce useful information without highly
specialized know-how or complicated interpretation are therefore
highly desirable.

Here we examine how
the sex chromosome system of a species can be identified from sequencing data
alone, and how a straightforward analysis based on sequencing coverage
can reveal the individual scaffolds that constitute the sex
chromosomes.
We target the minke whale genome and use
published data from several sequencing projects to demonstrate how this
method can be applied in practice.  The minke whale consists of two
species, the Antarctic \emph{Balaenoptera bonarensis}, and the common
minke whale \emph{Balaenoptera acutorostrata}, with two major
subspecies, the Atlantic \emph{B.~a.~acutorostrata} and Pacific minke
whale \emph{B.~a.~scammoni} \cite{rice1998marine}. 

\subsection{Sex chromosomes}
\label{sec:org5b2ba5b}

Sexually reproducing eukaryotic organisms are normally diploid, with
most of the chromosomes (the autosomes) occurring in somatic cells in
two copies, one copy inherited from each parent.  Sex
chromosomes (the allosomes) constitute the major exception to this rule.
In most cases, one sex is \emph{homogametic}, having two copies of one sex
chromosome, while the other sex is \emph{heterogametic}, having one copy of
each of the sex chromosomes.   In the latter case, the sex chromosomes
may be homomorphic or heteromorphic in variable degree depending on
the divergence between the sex chromosomes \cite{wright2016make}.
In mammals, for instance, the male is
heterogametic, and the chromosomes are referred to as X and Y,
while for many birds and reptiles, the female is heterogametic, and
the sex chromosomes are referred to as Z and W.
Although the typical case is that heterogametes have a single chromosome of each type,
some species lack a Y or W chromosome (then referred to as X0 or
Z0 systems), and several examples of species with multiple sex
chromosomes exist \cite{white1940origin,rens2007multiple}.

Often, the sex chromosomes which are present only in the
heterogametic sex (i.e. the Y or W chromosomes) are smaller and have
eroded to contain few genes compared to the corresponding sex
chromosome which is found in both sexes \cite{kaiser2010evolution}.
An obvious consequence of hemizygosity is a difference in gene
dosage which is often, but not always, compensated for in terms of
gene expression levels \cite{kaiser2010evolution}.

Local mutations can modify genes and thus affect the fitness of an
individual, but the evolution of chromosomes is also governed by local or genome-wide
rearrangements like insertions, duplications, deletions, inversions,
and translocations. Recombination and these genomic rearrangements can accelerate
adaptation by breaking up linkage between advantageous and
disadvantageous alleles.
Correspondingly, genetic linkage of sites under selection can hamper
adaptation by a combination of background selection (loss of non
deleterious variation at a locus due to negative selection against a
linked locus), Müllers ratchet (incremental accumulation of
deleterious mutations) or selective sweeps (loss of genetic variation
and accumulation of mutations in a region linked to a
locus under positive selection) collectively referred to as the
Robertson-Hill effect \cite{kaiser2010evolution,hill1966effect,eichler2003structural,comeron2008hill,presgraves2005recombination,barton2010genetic}.

Sex chromosomes play an important role in studying the above
since the adaptive
responsiveness of sex chromosomes is different to that of
autosomes due to their haploid state in the heterogametic sex and the
inherent smaller chromosome population size. As adaptability of a
trait varies with chromosomal location(s) of the underlying gene(s) it
is not surprising that the limited empirical data available appear to
indicate a non-random distribution of genes in the genome in general
and on sex chromosomes in particular
\cite{kaiser2010evolution,magnusson2012demasculinization,jaquiery2013masculinization}. Deciphering
the relative importance of the selective mechanisms, corroborating existing
knowledge on the evolution of sex chromosomes, and also
deciphering mechanisms compensating for gene dosage differences
between sexes all depend on increased numbers of sex chromosome datasets
available.  With the assumption that sequence coverage is
proportional to the amount of material used to construct the
sequencing libraries, the strategy used here can increase the number of sex
chromosome datasets by identifying heteromorphic regions of sex
chromosomes directly from the sequencing coverage. 

\subsection{Sequencing technology challenges}
\label{sec:orgbf381fb}

While the goal of genome sequencing often is to reconstruct the complete
chromosomes, current sequencing technology only
gives us shorter fragments. Illumina HiSeq instruments, for instance, typically produces
reads of 100 to 250 base pairs. In the ideal case, each location in
the chromosome would be equally likely to produce a sequence fragment.
This would give the sequencing coverage a
Poisson distribution.  For example, with 100 bp reads and
50x coverage, a 10 Kbp contig would then be expected to have about 5000
reads covering it, with a standard deviation of about 70.  A similar
haploid contig should then have an expected number of reads of about
2500 and a standard deviation of 50.  Thus from theory, these distributions
are separated by more than 40 standard deviations, and we can classify
contigs into one or the other with high accuracy.
In practice, the standard deviation is much larger, and in the
following, we will briefly look at some of the reasons for this.

The sequencing process itself has inherent coverage biases.  One well
known bias is tied to the AT/CG ratio of the sequences, with different
technologies responding differently to this ratio
(e.g., \cite{quail2012tale,dohm2008substantial}).
Another well-known if poorly understood cause for coverage bias is
artificial duplicates, where the same DNA fragment results in many
identical reads or read pairs (e.g., \cite{balzer2013filtering}).
Less relevant for genomic sequencing, the use of random hexamer
primers in RNA sequencing brings a marked bias in starting
points for sequences \cite{hansen2010biases}.

Coverage analysis relies on a good reference genome sequence, but this
quality is often hampered by the characteristics of the
specific genome.  One common obstacle is the repeat structure of the
genome, where various kinds of repeats (e.g., ALU LINE and SINE repeats and transposable
elements, ribosomal genes, and low complexity regions like di-
and tri-nucleotide repeats or microsatellites) mean that many genomic
regions can be difficult to distinguish from each other.  This can be
addressed by using long range information, for instance from mate-pair
sequencing, to bridge such regions, but the result is often scaffolds
with large regions of unknown sequence (by convention represented by
Ns).  The result is twofold, repeats are often collapsed in the
genome reference sequence, and these regions tend to see extremely
high coverage of sequencing data, and conversely, regions of Ns will
fail to map any reads, giving an apparent coverage of zero.
In addition, some species exhibit larger genomic variation than others, and many fish,
including the zebrafish \emph{(Danio rerio)} model organism, have proved challenging
\cite{brown2012extensive}.  This means that even when a high quality genomic reference can be produced,
it may not take into account genomic variation between individuals,
and as such, not be representative for the species.

In order to analyze coverage, we need to align (or map) sequencing
reads back to the reference genome.  Genomic features like repeats can
cause difficulties in the mapping, even when the reference sequence is
of high quality and correctly resolves these structures, and different
short read aligners (e.g. bwa \cite{li2009fast} and bowtie
\cite{langmead2012fast}) can have different strategies for dealing with
ambiguously aligned reads and short indels and other variations.

\subsection{Coverage based analysis}
\label{sec:org7defc0e}

Despite the difficulties outlined above, sequence coverage statistics
convey important information, and is used to aid many types of
analysis; e.g., sequencing error correction
\cite{simpson2012efficient,liu2013musket}, genome assembly 
\cite{chen2014tigra}, masking transcriptomic repeats \cite{malde2006rbr}, and analyzing
metagenomic data \cite{brown2012reference}, and is used directly to
aid statistical analysis (e.g., \cite{malde2014estimating}).

With sufficiently long
genomic scaffolds and sequencing data with a sufficiently high
coverage, it is possible to classify the ploidy of each scaffold in
an individual from the sequencing coverage. This approach was
used to successfully identify Y- and W-chromosomal genes in chicken
\cite{chen2012identification} and
mosquito \cite{hall2013six}, and to classify scaffolds in fruit fly
\cite{carvalho2013efficient}. Similar methods have since been applied 
to other cases (see \cite{tomaszkiewicz2017and} for a review).  Coverage
analysis has also been used to reveal important information about the
evolution and origin of chromosomes
\cite{fraisse2017deep,vicoso2013comparative,vicoso2013reversal}
In practice, it is possible to both identify the type of sex
chromosome arrangement, and to classify each individual scaffold by
the type of chromosome it belongs to.  The overall expected diploid
coverage (2n) is estimated for each sequencing library, and it is then
compared to the coverage for individual scaffolds using Table
\ref{tab:orgcdc00e8}.

\begin{table}[htbp]
\centering
\begin{tabular}{llll}
System & Male & Female & Class\\
\hline
 & 2n & 2n & autosomal\\
\hline
XY or X0 & 1n & 2n & X\\
XY & 1n & 0 & Y\\
\hline
ZW or Z0 & 2n & 1n & Z\\
ZW & 0 & 1n & W\\
\end{tabular}
\caption{\label{tab:orgcdc00e8}\small
\small Classification table for scaffolds based on sequencing coverage.  By classifying a scaffold as having 0, 1n or 2n coverage, where 2n represents the coverage expected from diploid autosomes, we can classify each scaffold by the type of chromosome.  From the classification of all scaffolds, we can infer the sex chromosome system for the species.}
\end{table}

Recently, a high quality genome assembly of a male Pacific
\emph{(B.~a.~scammoni)} minke whale
that was published by Yim et al \cite{yim2014minke}.  The assembly
consists of 104'326 scaffolds, but many are quite short, and only 1003
scaffolds are over 10'000 bp long.  Shortly after the publication, the
genome and (associated sequencing data) of the Antarctic minke
(\emph{B.~bonaerensis}) was published \cite{kishida2015aquatic}.
In addition, we had available full-genome sequencing data of an
Atlantic male minke whale (\emph{B.~a.~acutorostrata})
\cite{malde2017whole}.  Here, we apply coverage statistics to the
analysis of these data, and contrast our findings to previous
knowledge about the minke whale sex chromosomes.  We show that the
method also works across species as long as they are sufficiently
closely related.  Using this as an example, we argue that this
procedure, with little cost and effort, reveals more complete and
detailed information than what is typically reported in current
sequencing projects.

\section{Materials and methods}
\label{sec:org5b0ae56}

The minke whale reference genome was downloaded from the \texttt{bioftp.org} website.
Sequencing data for the two previously published genome projects
\cite{yim2014minke,kishida2015aquatic}, listed in Table \ref{tab:orga390d1f} were
downloaded as FASTQ-formatted text files from the EBI FTP site.

\begin{table}[htbp]
\centering
\begin{tabular}{lllrl}
Species & Sex & Type & Cov & Comment\\
\hline
B.a.scammoni & M & ind, several libs & 1.26 & Yim et al. \cite{yim2014minke}\\
B.a.scammoni & F & 3x ind runs & 1.1 & Yim et al. \cite{yim2014minke}\\
B.a.acutorostrata & M & ind, 2 runs & 0.28 & Malde et al. \cite{malde2017whole}\\
B.bonaerensis & ? & ind, 1 run & 0.98 & Kishida et al. \cite{kishida2015aquatic}\\
\hline
\end{tabular}
\caption{\label{tab:orga390d1f} \small Overview of minke whale data sets used.}
\end{table}

Sequences were mapped to the reference genome using \texttt{bwa}
version  0.7.10-r789 \cite{li2009fast} using the \texttt{mem} alignment method.  The resulting
SAM files were then compressed and sorted, and mapping statistics
extracted with \texttt{samtools} version 0.1.19-96b5f2294
\cite{li2009sequence}.  The list of files is given in Table \ref{tab:orged738f0}.

\begin{table}[htbp]
\centering
\begin{tabular}{llll}
SRA ID & Specimen & Size & Mapped\\
\hline
SRR4011108 & Atlantic \emph{(B.a.acutorostrata)} male \cite{malde2017whole} & 42G & 91.3\%\\
SRR4011112 &  & 42G & 90.5\%\\
\hline
DRR014695 & Antarctic \emph{(B.bonaerensis)} \cite{kishida2015aquatic} & 315G & 89.5\%\\
\hline
SRR893003 & Pacific \emph{(B.a.scammoni)} male \cite{yim2014minke} & 73G & 92.1\%\\
SRR896642 &  & 27G & 90.6\%\\
SRR901891 &  & 67G & 92.0\%\\
SRR908213 &  & 50G & 96.8\%\\
SRR914419 &  & 30G & 96.9\%\\
SRR915980 &  & 24G & 93.6\%\\
SRR917970 &  & 9G & 96.3\%\\
\hline
SRR924087 & Pacific \emph{(B.a.scammoni)} female 1 \cite{yim2014minke} & 103G & 99.3\%\\
SRR924103 & Pacific \emph{(B.a.scammoni)} female 2 \cite{yim2014minke} & 93G & 99.4\%\\
SRR926179 & Pacific \emph{(B.a.scammoni)} female 3 \cite{yim2014minke} & 85G & 99.3\%\\
\end{tabular}
\caption{\label{tab:orged738f0}\small
List of datasets mapped to the reference genome.  All datasets map well to the Pacific reference, indicating that the species and subspecies are closely related.  The very high mapping rate for the Pacific females likely means the sequences were filtered before submission to SRA.}
\end{table}

We wrote a small program to count the number of N characters in
each scaffold, and the mapping statistics reported by \texttt{samtools} were
then normalized to the scaffold length subtracted the number of Ns.

Bérubé and Palsboll \cite{berube1996identification} identified PCR
primers (replicated here in Table \ref{tab:org5f5d184}) that provided an
effective sex assay for minke whales.  These primers give two different PCR products in
males (212 and 245 bp in size), but only a single (245 bp) product in
females, and therefore distinguish between the two sexes.
The sex marker primer sequences were matched against the minke genome
scaffolds using \texttt{blastn} with an E-value threshold of 0.1, and otherwise
using default parameters.

\begin{table}[htbp]
\centering
\begin{tabular}{ll}
Name & Sequence\\
\hline
ZFYX-1 & \texttt{ATAGGTCTGCAGACTCTTCTA}\\
ZFYX-2 & \texttt{ATTACATGTCGTTTCAAATCA}\\
ZFYX-3 & \texttt{CACTTATGGGGGTAGTCCTTT}\\
\end{tabular}
\caption{\label{tab:org5f5d184}\small
PCR primer sequences for the minke sex marker, as reported in previous studies \cite{berube1996identification}.}
\end{table}

\section{Results and discussion}
\label{sec:orgff26a9b}

\subsection{PCR-based sex markers}
\label{sec:org44bad39}

The PCR primers identified by Bérubé and Palsboll
\cite{berube1996identification} were matched against the genome assembly
using BLAST, resulting in the matches given
in Table \ref{tab:orga48a2e7}.  We see that scaffold351 matches ZFYX-3 in the
forward direction, and ZFYX-1 in the reverse direction, 243 bp further
downstream.  Similarly scaffold380 matches ZFYX-2 and ZFYX-3 in the
forward direction, and ZFYX-1 in the reverse direction, with distances
of 243 bp and 210 bp.  These matches are consistent with the products
described by Bérubé and Palsboll, and we conclude that scaffold351 (of
length 468 Kbp) originates from the X chromosome, while
scaffold380 (274 Kbp) is from the Y chromosome.

\begin{table}[htbp]
\centering
\begin{tabular}{llrrrrrrrrrr}
Query & Target & Id & len & subst & gaps & QStart & QEnd & TStart & TEnd & Eval & Score\\
\hline
ZFYX-1 & scaffold3112.1 & 100.00 & 21 & 0 & 0 & 1 & 21 & 61 & 41 & 0.002 & 42.1\\
ZFYX-3 & scaffold351 & 100.00 & 21 & 0 & 0 & 1 & 21 & 86228 & 86248 & 0.002 & 42.1\\
ZFYX-1 & scaffold351 & 100.00 & 21 & 0 & 0 & 1 & 21 & 86471 & 86451 & 0.002 & 42.1\\
ZFYX-3 & scaffold380 & 100.00 & 20 & 0 & 0 & 1 & 20 & 44203 & 44222 & 0.008 & 40.1\\
ZFYX-2 & scaffold380 & 100.00 & 21 & 0 & 0 & 1 & 21 & 44236 & 44256 & 0.002 & 42.1\\
ZFYX-1 & scaffold380 & 100.00 & 21 & 0 & 0 & 1 & 21 & 44446 & 44426 & 0.002 & 42.1\\
ZFYX-3 & scaffold8069.1 & 100.00 & 20 & 0 & 0 & 1 & 20 & 20 & 1 & 0.008 & 40.1\\
\end{tabular}
\caption{\label{tab:orga48a2e7}\small
List of BLAST matches in the minke genome assembly for the sex marker primer sequences.  This identifies scaffold351 and scaffold380 as belonging to the X and Y chromosomes, respectively.}
\end{table}

\subsection{Alignment-based identification of sex chromosomes}
\label{sec:orge51b02d}

As part of the Pacific minke genome sequencing project, Yim et al
\cite{yim2014minke} identified eight putative
scaffolds as sex chromosomes by using LASTZ \cite{harris2007improved} to match them against bovine sex
chromosomes.  According to their Supplementary Table 16, the matches
covered from 56\% to 72\% of the contigs.  The length of these contigs
totaled 807 Kbp, and contained 11 identified genes.

These pre-existing resources identify ten
scaffolds of a total length of 1549 Kbp of scaffolds classified as
putative sex chromosomes.  In contrast, the human genome is similar in
size to the minke whale, and has an X chromosome size of 155 Mbp,
almost precisely a hundred times larger than these scaffolds.  The
human X chromosome is estimated to contain 1098 genes
\cite{ross2005dna}.  Chromosome sizes vary between species, but for
mammals where the genome assemblies have sufficient quality for NCBI
Genomes to list individual chromosome sizes (Table \ref{tab:org0d0cd46}), X
chromosomes range from 127 (cat) to 171 (mouse) Mbp.  The Y
chromosomes vary more in size, and the size estimates are less
reliable due to repeats, low complexity, and heterochromatic sequence,
but they are normally much smaller than
X chromosomes.  Unless the minke whale is very atypical,
these earlier works have only identified a small fraction of the sex
chromosomes, perhaps as little as one percent.

\begin{table}[htbp]
\centering
\begin{tabular}{llrl}
Scientific name & Vernacular & X & Y\\
\hline
Homo sapiens & Human & 156 & 57\\
Bos taurus & Cow & 148 & -\\
Equus caballus & Horse & 124 & -\\
Felis catus & Cat & 127 & -\\
Mus musculus & Mouse & 171 & 91\\
Rattus norwegicus & Rat & 159 & 3\\
\end{tabular}
\caption{\label{tab:org0d0cd46}\small
Sizes of X and Y chromosomes (in megabases) for various species as reported by the NCBI Genomes web site.}
\end{table}

\subsection{Sequence coverage for the Pacific minke whale}
\label{sec:org19d40f3}

Examining how the sequencing data maps back to the genome
assembly by plotting the number of mapped sequences from the male
specimen on each contig (or scaffold)
against the contig length (Fig 1), it is 
clear that most of the data points cluster along one of two
horizontal axes: one centered around y=1.2 and one at y=0.6.  The
obvious hypothesis is that these correspond to autosomal chromosomes
and allosomal (sex chromosomes) respectively, and we will investigate
this further below.

\begin{figure*}[htbp]
\centering
\includegraphics[width=.9\linewidth]{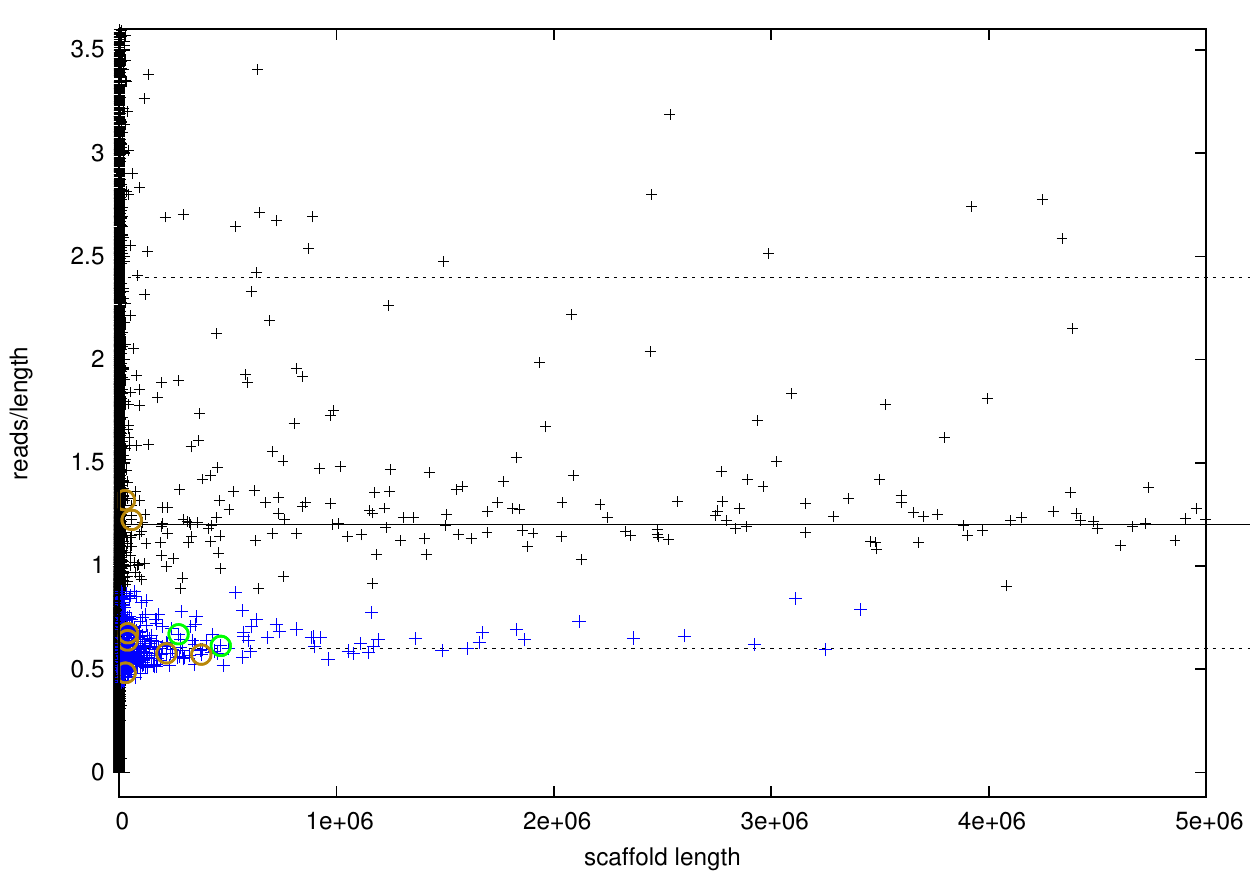}
\caption{\label{fig:orgb5e30c2}\small
\textbf{Sequencing coverage by contig length for the male Pacific minke whale used sequence by Yim et al \cite{yim2014minke}.}  Each point represents one genomic scaffold, scaffolds classified as belonging to sex chromosomes by Yim et al are circled in orange, scaffolds carrying the sex markers \cite{berube1996identification} are circled in green.  Scaffolds classified as sex chromosomes from sequence coverage are marked in blue.}
\end{figure*}

In Fig 1, as well as in the following diagrams, contigs with coverage 
between 0.28 and 0.88 in the Pacific male have been marked in blue.
In addition, the eight putative sex chromosome contigs identified by Yim
et al \cite{yim2014minke} have been circled in orange, and the two contigs matching the
PCR primers of \cite{berube1996identification} have been circled in green.

From Fig 1, it is apparent that not all of the putative sex
chromosome scaffolds (i.e., orange circles) appear to have a coverage
consistent with allosomal scaffolds.  We will return to this point
below.
We see that for short contigs, the data become quite noisy.
It is clear from the theoretical Poisson distribution that variance is
a function of length, so short contigs (with correspondingly few
mapped reads) will tend to have a high variance.  But more
importantly, short contigs are often a result of problems in the
assembly process.  For instance, repeats or low complexity regions are
difficult for the assembler software to resolve, leading to
ambiguities in the sequence that forces it to terminate contigs
early.

In addition to the reference individual, Yim et al \cite{yim2014minke}
sequenced three female specimens.  In order to
corroborate our hypothesis, we can contrast the results from the male
with a similar plot of the data from three female specimens, from the same project.

Here (Fig 2) we see that the data is centered along a single axis, 
corresponding to a coverage of approximately 1.1.  Although there are
many quite short contigs with highly variable coverage, there appears to be a small cluster of
slightly longer contigs with low coverage in the bottom left
corner.  No such cluster is apparent in the data from the male, and it
is tempting to classify these as belonging to the Y-chromosome, with
the apparent coverage resulting from sequence similarity between the
X and Y chromosomes.  This
conjecture is supported by the Y-marker scaffold being placed clearly
in this group, while the X-marker scaffold has coverage consistent
with the autosomal scaffolds.

\begin{figure*}[htbp]
\centering
\includegraphics[width=.9\linewidth]{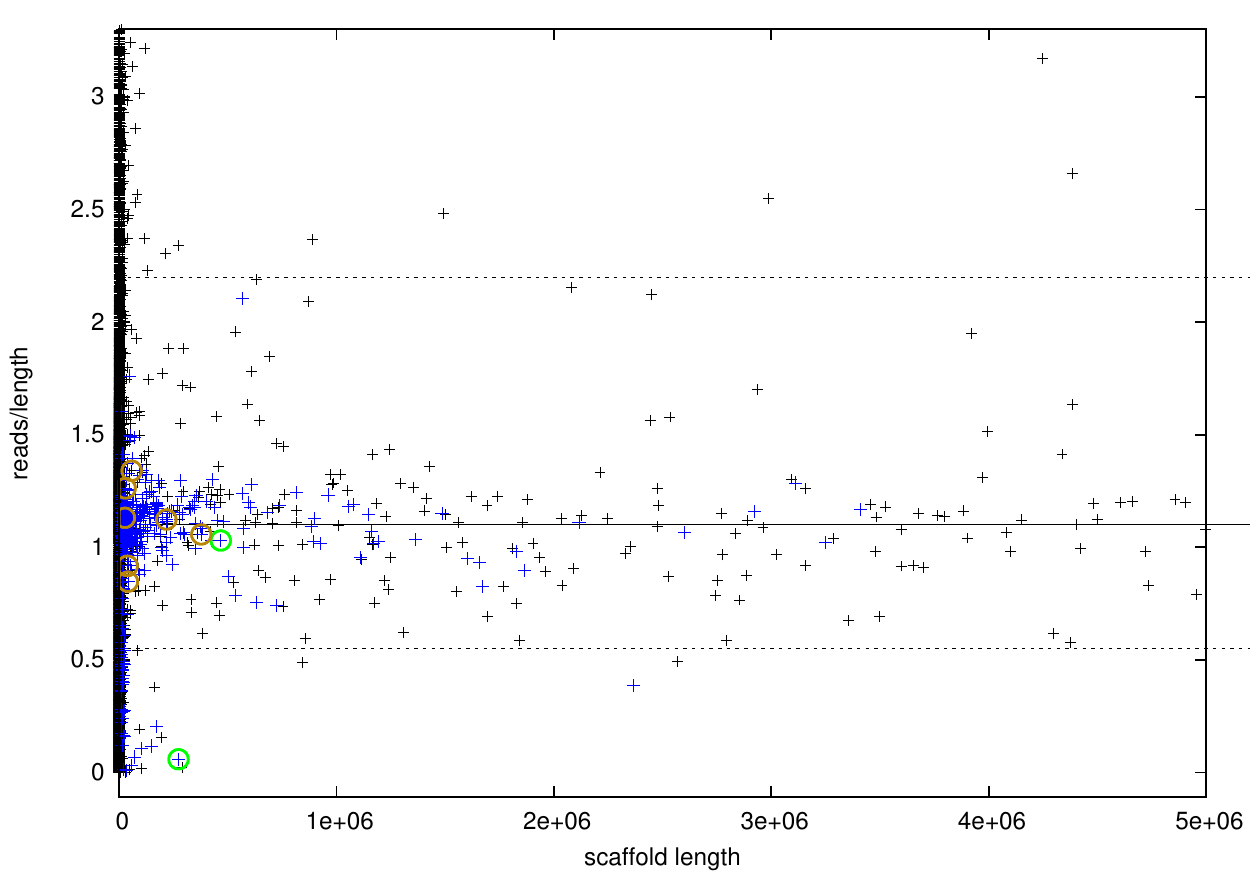}
\caption{\label{fig:orgc374aa4}\small
\textbf{Sequencing coverage by contig length for three female Pacific minke whales \cite{yim2014minke}.}  Each point represents one genomic scaffold, scaffolds classified as belonging to sex chromosomes by Yim et al are circled in orange, scaffolds carrying the sex markers \cite{berube1996identification} are circled in green.  Scaffolds classified as sex chromosomes from sequence coverage are marked in blue.}
\end{figure*}

If we examine the correlation between coverage in the Pacific male
with the coverage in the Pacific female (Fig 3), we see that 
the coverage data forms two clear clusters, one representing sex
chromosomes with half the normal coverage in the male (as before, blue color
indicates coverage between 0.28 and 0.88 in the Pacific male data),
and another representing autosomes in red.
For clarity, only contigs longer than 10 Kbp are highlighted in color,
shorter contigs are shown in gray. Short contigs often arise from
sequencing errors, and therefore tend to be noisier
and have more variable coverage.  We observe that most of the
putative sex chromosome scaffolds are placed as expected.  Also, the
Y-marker scaffold is placed with close to zero coverage in the female.
However, there isn't a clear cluster surrounding this scaffold, but
rather a set of scaffolds with haploid coverage in male, but a range
of coverages stretching from zero to diploid coverage in 
females.  This is likely due to similarity between homologous regions
between the X and Y chromosomes, causing some X chromosomal reads to
map to Y chromosomal scaffolds.

\begin{figure*}[htbp]
\centering
\includegraphics[width=.9\linewidth]{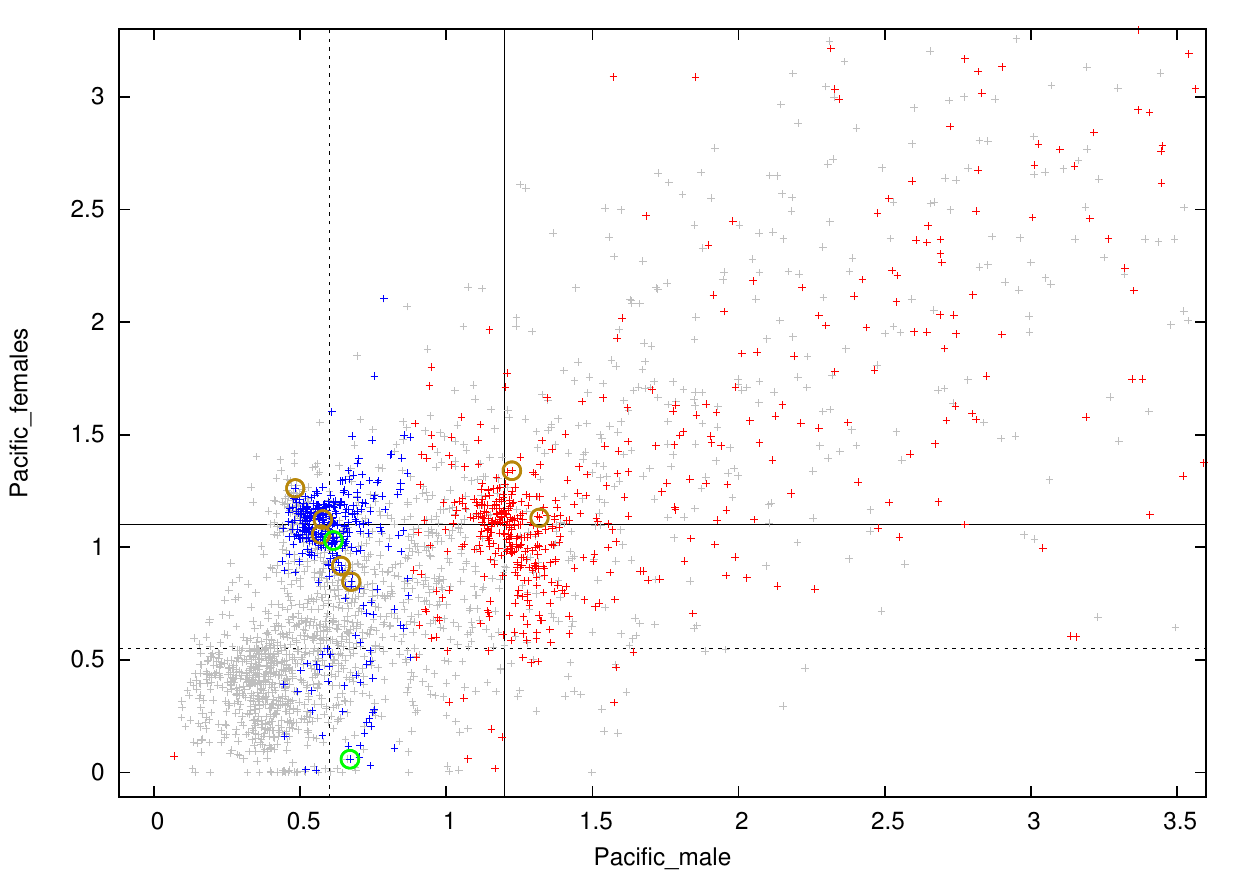}
\caption{\label{fig:org249fa92}\small
\textbf{Comparison of sequencing coverage for a Pacific male and three Pacific females \cite{yim2014minke}.}  Each point represent a scaffold, blue points correspond to scaffolds classified as belonging to sex chromosomes, red points represent autosomal scaffolds, while grey points represent very short scaffolds.  Scaffolds classified as belonging to sex chromosomes by Yim et al are circled in orange, scaffolds carrying the sex markers \cite{berube1996identification} are circled in green.}
\end{figure*}

\subsection{The Atlantic subspecies}
\label{sec:org2e0dc48}

The Atlantic minke whale (\emph{B.~acutorostrata acutorostrata}) is, like
the Pacific minke (\emph{B.~acutorostrata scammoni}), considered a
subspecies of the common minke whale (\emph{B.~acutorostrata}).  Although there are clear genetic
differences between them \cite{pastene2007radiation}, we can expect
the two subspecies to be sufficiently closely related that sequences from
the Atlantic minke can be mapped to the Pacific reference assembly with little error.

In Fig 4, we see that this indeed seems to be the case.  Here, we 
have taken sequences from a male Atlantic specimen, and mapped them to
the (\emph{B.~a.~scammoni}) reference genome.  We see that the emerging picture is
consistent with the data from the Pacific male, clearly identifying
the specimen as a male.  One exception is the two putative sex
chromosomal scaffolds which exhibited autosomal coverage in the
Pacific specimen.  In the Atlantic male, one of them has coverage
consistent with sex chromosomes, while the other has in fact increased
coverage.  This is perhaps even clearer from Fig 5, which 
displays the correlation between coverage in the Pacific male and the
Atlantic male specimens.  Most likely, these scaffolds are repeats
that vary in copy number between individuals.  Apart from these
scaffolds, the data groups clearly into two clusters consistent with
autosomal and allosomal chromosomes.

\begin{figure*}[htbp]
\centering
\includegraphics[width=.9\linewidth]{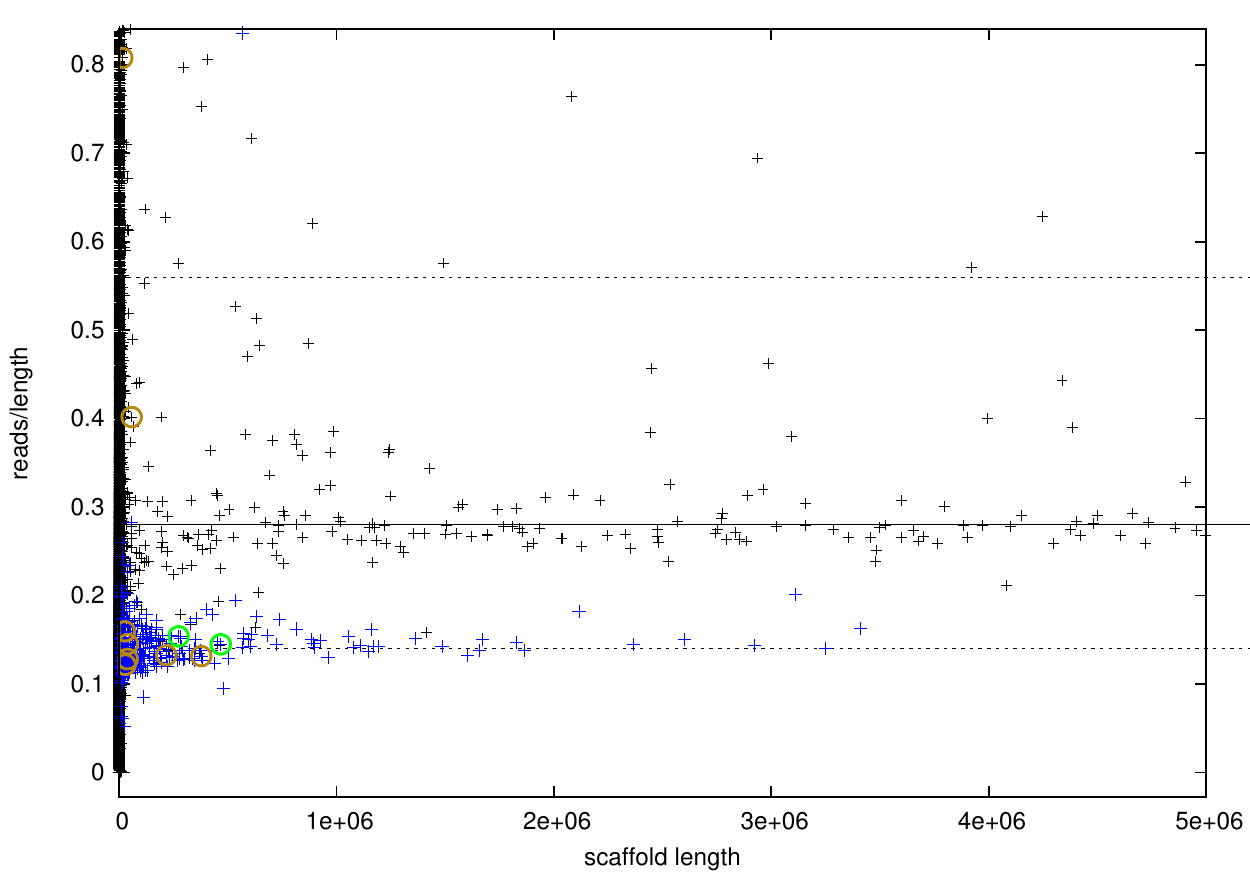}
\caption{\label{fig:org9ddf074}\small
\textbf{Sequencing coverage by contig length for a male Atlantic minke whale.}  Each point represents one genomic scaffold, scaffolds classified as belonging to sex chromosomes by Yim et al are circled in orange, scaffolds carrying the sex markers \cite{berube1996identification} are circled in green.  Scaffolds classified as sex chromosomes from sequence coverage are marked in blue.}
\end{figure*}

\begin{figure*}[htbp]
\centering
\includegraphics[width=.9\linewidth]{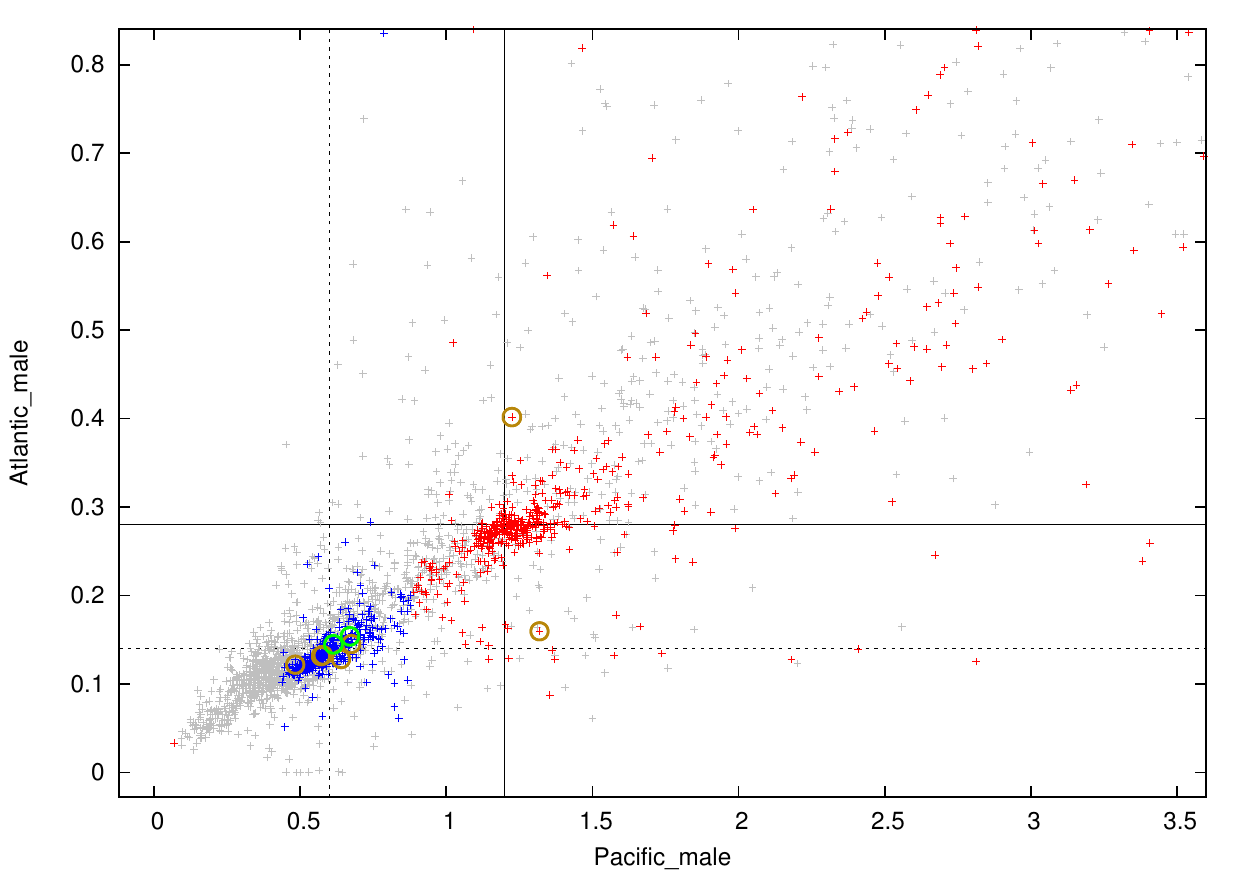}
\caption{\label{fig:org5017697}\small
\textbf{Comparison of sequencing coverage for an Atlantic male and a Pacific male \cite{yim2014minke}.}  Each point represent a scaffold, blue points correspond to scaffolds classified as belonging to sex chromosomes, red points represent autosomal scaffolds, while grey points represent very short scaffolds.  Scaffolds classified as belonging to sex chromosomes by Yim et al are circled in orange, scaffolds carrying the sex markers \cite{berube1996identification} are circled in green.}
\end{figure*}

\subsection{The Antarctic minke whale}
\label{sec:org41b310d}

In a recent study, Kishida et al \cite{kishida2015aquatic} used DNA sequencing of an
Antarctic minke whale (\emph{B.~bonaerensis}) to study the genetic effects of aquatic adaption
of mammals.  The Antarctic minke is more distantly related, and
it is considered a separate species \cite{arnason1993cetacean}.  The individual studied
by Kishida et al. \cite{kishida2015aquatic}, was of undetermined sex, but by mapping the data
and comparing Fig 6 to the figures above, it becomes clear that 
this is a female specimen.

\begin{figure*}[htbp]
\centering
\includegraphics[width=.9\linewidth]{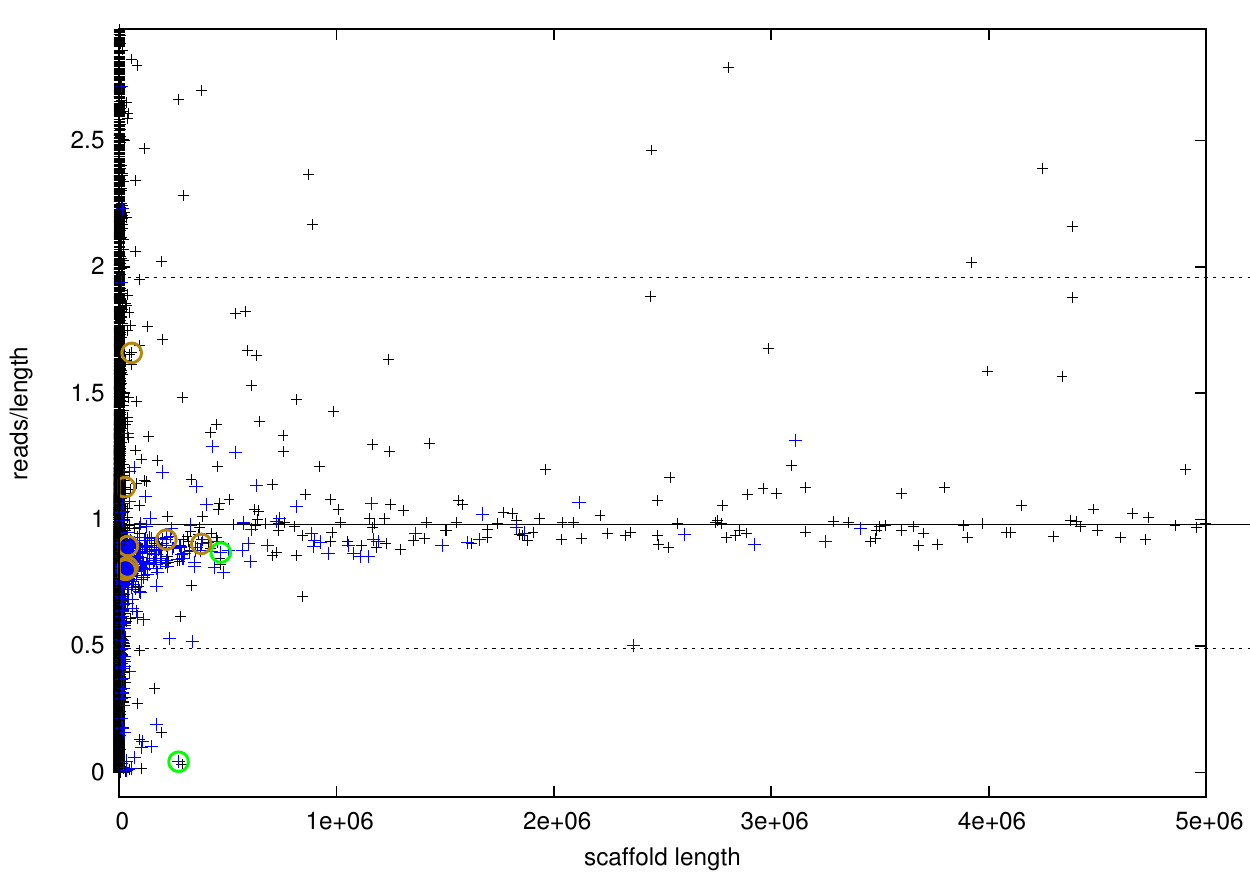}
\caption{\label{fig:org97502bc}\small
Sequencing coverage by contig length for an Antarctic minke whale \cite{kishida2015aquatic}. The sex is not given, but it is clear from the sequencing coverage that the specimen is female. Each point represents one genomic scaffold, scaffolds classified as belonging to sex chromosomes by Yim et al are circled in orange, scaffolds carrying the sex markers \cite{berube1996identification} are circled in green.  Scaffolds classified as sex chromosomes from sequence coverage are marked in blue.}
\end{figure*}

From comparing coverage of Pacific male to the Antarctic female (Fig 7) 
we see a similar picture, but note that coverage for the X
cluster in the female is lower than for the autosomal cluster.  We
speculate that this could be
an artifact of rapid evolution of haploid chromosomes,
leading to somewhat fewer sequences matching it.

\begin{figure*}[htbp]
\centering
\includegraphics[width=.9\linewidth]{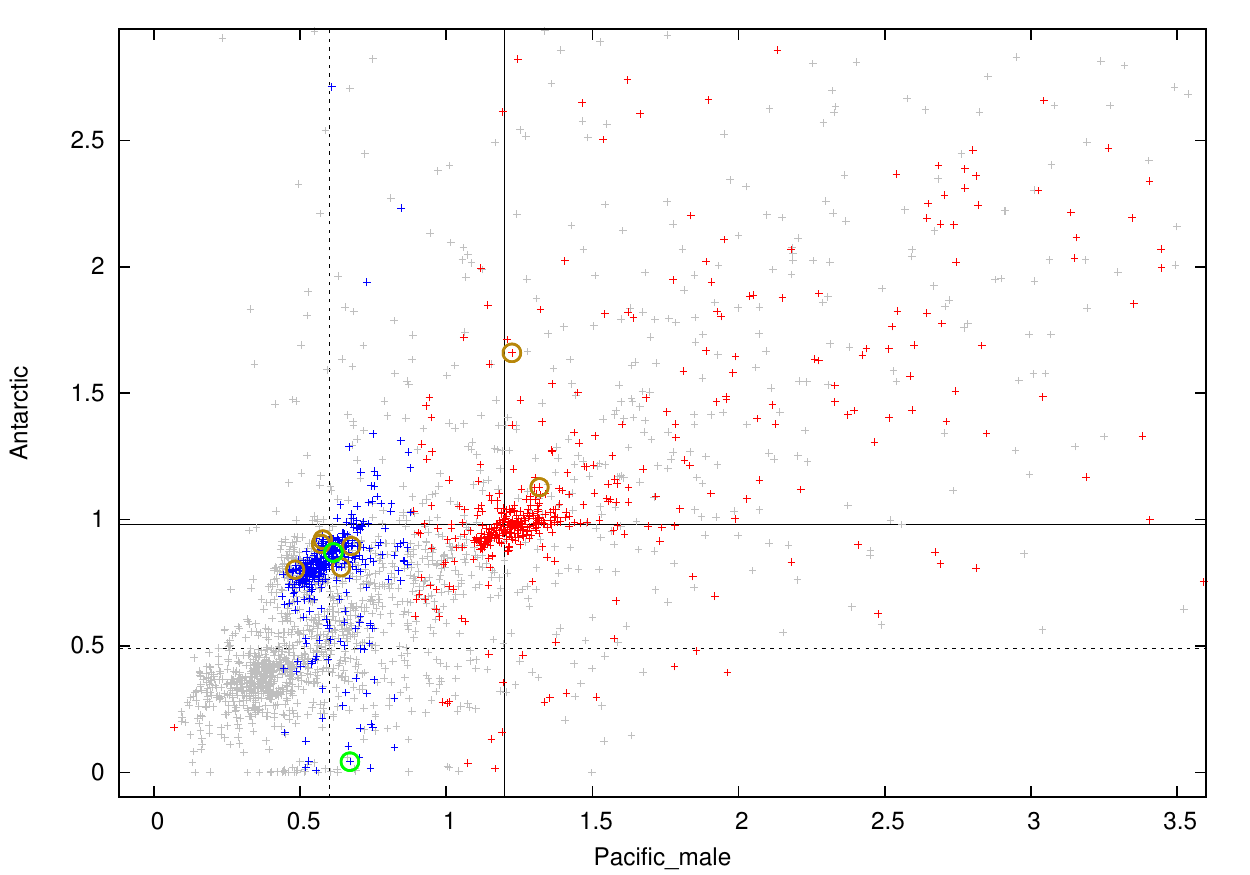}
\caption{\label{fig:orgffa1bcd}\small
\textbf{Comparison of coverage for a Pacific male \cite{yim2014minke} to an Antarctic specimen \cite{kishida2015aquatic}.}  Each point represent a scaffold, blue points correspond to scaffolds classified as belonging to sex chromosomes, red points represent autosomal scaffolds, while grey points represent very short scaffolds.  Scaffolds classified as belonging to sex chromosomes by Yim et al are circled in orange, scaffolds carrying the sex markers \cite{berube1996identification} are circled in green.}
\end{figure*}

\subsection{Sorting out contigs}
\label{sec:org8b08b4b}

In Table \ref{tab:org19d0b22} we look in more detail at the coverage scaffolds
identified as sex chromosomes by \cite{yim2014minke}, and the two
scaffolds matching the sex markers from \cite{berube1996identification}.
Of the eight scaffolds classified as X chromosomal by Yim et al, five have coverage
consistent with X-chromosomal scaffolds both in males and females.
The scaffolds identified by the markers fit the expected coverage from
X and Y almost perfectly.

\begin{table}[htbp]
\centering
\begin{tabular}{llrrrr}
NCBI Accession & Scaffold & PM & PF & AM & Ant\\
\hline
\texttt{NW\_006732179.1} & scaffold725 & 0.403 & 1.147 & 0.435 & 0.819\\
\texttt{NW\_006728548.1} & scaffold398 & 0.481 & 1.022 & 0.473 & 0.940\\
\textbf{\texttt{NW\_006733712.1}} & \textbf{scaffold863} & \textbf{4.161} & \textbf{7.436} & \textbf{2.890} & \textbf{5.457}\\
\texttt{NW\_006731312.1} & scaffold647 & 0.567 & 0.770 & 0.519 & 0.914\\
\textbf{\texttt{NW\_006730490.1}} & \textbf{scaffold573} & \textbf{1.020} & \textbf{1.218} & \textbf{1.435} & \textbf{1.694}\\
\texttt{NW\_006728042.1} & scaffold352 & 0.476 & 0.961 & 0.469 & 0.923\\
\textbf{\texttt{NW\_006732401.1}} & \textbf{scaffold745} & \textbf{1.099} & \textbf{1.029} & \textbf{0.570} & \textbf{1.152}\\
\texttt{NW\_006731168.1} & scaffold634 & 0.533 & 0.834 & 0.458 & 0.829\\
\hline
\texttt{NW\_006728031.1} & scaffold351 & 0.511 & 0.937 & 0.518 & 0.888\\
\texttt{NW\_006728349.1} & scaffold380 & 0.558 & 0.054 & 0.550 & 0.044\\
\end{tabular}
\caption{\label{tab:org19d0b22}\small
Scaffolds and average coverage in the different data sets.  The first eight were identified by Yim et al., using alignment, the last two were found to contain the sex marker primer sites identified by Bérubé and Palsboll \cite{berube1996identification}.  Coverage is normalized, so that 1.0 corresponds to expected diploid coverage.  The specimens are Pacific male (PM), three Pacific females (PF), Atlantic male (AM), and an Antarctic specimen of undetermined sex (Ant).}
\end{table}

Of the remaining three scaffolds, we see that scaffold573 appears to have roughly normal
diploid coverage,  the slightly higher coverages for Atlantic and
Antarctic minke may be indicative of repeat structures.
Scaffold745 is interesting, as it appears diploid in Pacific minke, but is close to haploid
in Atlantic minke.  That the observed coverage deviates
from the expected coverage for an allosome does not disprove the claim of their
X-chromosomal origin, but suggests that a closer investigation should be
performed before their classification can be finally determined.
Scaffold863 has about 4x the expected (diploid)
coverage in Pacific male, and close to twice that in the Pacific
females.  Coverage is lower in Atlantic and Antarctic, but viewed
independently, this is consistent with an X-chromosomal scaffold.
It seems likely this scaffold is (or contains) a repeat, and that copy
number variation is responsible for the variation between the groups.

From coverage we estimate 18'907 scaffolds belong to the
allosomes, representing a total length of 97'232'273, or 85'666'973
not counting Ns.  Most of these scaffolds are short, and only 357 of
them are longer than 10 Kbp, totaling 91'066'100, or 79'974'798 bp
not counting Ns.  Thus, despite the large number of short scaffolds,
they constitute only 6.3\% of the putative allosomal sequences.

For reasons outlined above, segregating Y from XY is more difficult,
but if we take a coverage of 0.6 in the Pacific
female data as a threshold, we identify a subset of 5'318 scaffolds
representing 5'973'998 bp (5'410'860 bp without Ns), or 37 scaffolds
longer than 10 Kb representing 3'955'760 bp (3'602'523 bp without Ns).

\subsection{Genes on the sex chromosomes}
\label{sec:org7e54e8f}

Yim et al \cite{yim2014minke} identify 27109 genes in the complete genome (including 13 on the mitochondrial
genome, and 4084 pseudogenes).  Matching this list against the 357
scaffolds longer than 10 Kbp that were identified as sex chromosomes
from the coverage data, we identified a set of 1007 genes.  This is
summarized in Table \ref{tab:orgd5b4bc1}.  The sets are highly redundant, and only two
genes identified as X chromosomal by Yim et al., FAM123B
and ASB12, both from \texttt{scaffold573}, were not corroborated by the coverage analysis.

\begin{table}[htbp]
\centering
\begin{tabular}{lrr}
Method & scaffolds & genes\\
\hline
Markers & 2 & 18\\
Alignment & 8 & 13\\
Coverage & 357 & 1007\\
\hline
\end{tabular}
\caption{\label{tab:orgd5b4bc1}\small
Scaffolds over 10 Kbp that are considered sex chromosomal, based on the available information. "Alignment" refers to the scaffolds identified by Yim et al by alignment to the bovine sex chromosomes.}
\end{table}

\section{Conclusions}
\label{sec:orge2a097f}

Classifying scaffolds by coverage is straightforward and
works well.  Although multiple high-coverage libraries are desirable
(and often available), a single sequencing library from each sex can,
when mapped to a high quality genome assembly, be sufficient to
separate sex chromosomal sequences from autosomes.  Here we use an
Illumina HiSeq genomic shotgun library, but other technologies like RADseq
\cite{gamble2015restriction} and transcriptome sequencing
\cite{rovatsos2016mammalian} can also be used
to good effect.
In contrast, commonly used
methods based on homology tend to fail to correctly identify much of the sex chromosomal
sequences. 
With some care in
planning of experiments and sequencing, any genome project should be
able to integrate this analysis, and thereby identify a large fraction
of scaffolds representing sex chromosomes.  The sizes of the minke
whale allosomes are not known, but if they are close to the mammalian average, the 90 Mbp
of large scaffolds we have identified in this study would represent approximately 60\% of
the total. More importantly, repeats, pseudoautosomal regions, and low complexity sequence which
confounds this analyis are often found in the intergenic regions.  This
number is therefore a conservative estimate for sex chromosomal genes,
and the number of discovered genes is reasonable when compared to the
number of sex chromosomal genes in mammals.

Although we have here only applied this method to the minke whale, we
expect the method would work  well for other species with XY or ZW sex
chromosome systems.  There are, however, some important limitations.  In
particular, a high quality reference genome sequence is necessary,
where contigs are long enough to even out local variations in the
sequencing coverage.  In our analysis, we found that scaffolds longer than
10 Kbp were sufficient.  We further found that adjusting for Ns in the scaffolds improved the results.  The presence of repeats did not appear
to cause similar problems here, it is possible that for genome assemblies
where many repeat sequences are collapsed, this can skew coverage
statistics.  One should also be aware of other causes of
coverage variation. 
Many species have small haploid chromosomes called
B-chromosomes whose presence may vary between
individuals \cite{nla.cat-vn1871086}.  For species with a high degree of genomic variation
between individuals, more care may need to be taken in sampling and
sequencing for coverage analysis to be effective.
Sex chromosomes in particular often have heterochromatic
sequence, repeats, and homologous regions that may not be possible to
resolve using mapping coverage alone.  Sequence coverage, like 
many computational methods, thus serves as a starting point, and
results should be verified by independent, experimental methods like
PCR, qPCR or FISH.

\subsection{Estimation of expected coverage}
\label{sec:orgdff1585}

We do not here provide
a rigorous statistical analysis for estimating the average coverage.
Although it is common to discuss this in analytical (typically
assuming reads are Poisson distributed) or empirical terms, in our
experience (and as discussed above) sequencing data are fraught with
errors and biases.  To get a good fit in practice, some amount of
judgement must be applied.  In any case, we find the clustering is
apparent from the data, and thus we focus on the biological
implications.  For readers interested in pursuing a rigorous approach,
we here provide some pointers.

Starting with the assumption of coverage approximately following a
Poisson distribution, we run into the complication of
heteroskedasticity - specifically, that the variance in coverage over
a long scaffold will be lower than the variance over short scaffolds.
This variation is evident from the trumpet shape of the plots. To deal
with this, it is common to adjust the Poisson model with an exposure
variable, using the log size of the exposure (i.e., scaffold length)
as a parameter with a fixed coefficient of 1.

In reality, coverage is not Poisson distributed, and repeats, assembly
errors and sequencing errors and biases are likely to cause
overdispersion.  In this case, a negative binomial distribution is
likely to be a better approximation.

In the case of comparing a male to a female specimen, we would like to
model the data as coming from a mixture of distributions (see Table
\ref{tab:orgcdc00e8}) with different coefficients (corresponding to the relative
sizes of the autosome and sex chromosomes).  This is a good
application of the Expectation Maximization algorithm, which could be
used with either a Poisson distribution with exposure, or negative
binomials.

\subsection{Availability}
\label{sec:org361d40a}

All data used is publicly available as described in the relevant
publications \cite{yim2014minke,kishida2015aquatic,malde2017whole}.  The
scripts used in the analysis are released in the public domain and
available from GitHub (\url{https://github.com/ketil-malde/sexchrcov}).


\begin{thebibliography}{10}

\bibitem{yim2014minke}
Yim HS, Cho YS, Guang X, Kang SG, Jeong JY, Cha SS, et~al.
\newblock Minke whale genome and aquatic adaptation in cetaceans.
\newblock Nature genetics. 2014;46(1):88--92.


\bibitem{rice1998marine}
Rice DW
\newblock Marine mammals of the world.
\newblock Systematics and distribution. 1998.

\bibitem{wright2016make}
Wright AE, Dean R, Zimmer F, Mank JE.
\newblock How to make a sex chromosome.
\newblock Nature communications. 2016;7:12087.

\bibitem{white1940origin}
White MJ.
\newblock The origin and evolution of multiple sex-chromosome mechanisms.
\newblock Journal of Genetics. 1940;40(1):303--336.

\bibitem{rens2007multiple}
Rens W, O’Brien P, Grutzner F, Clarke O, Graphodatskaya D, Tsend-Ayush E,
  et~al.
\newblock The multiple sex chromosomes of platypus and echidna are not
  completely identical and several share homology with the avian Z.
\newblock Genome Biol. 2007;8(11):R243.

\bibitem{kaiser2010evolution}
Kaiser VB, Bachtrog D.
\newblock Evolution of sex chromosomes in insects.
\newblock Annual review of genetics. 2010;44:91.

\bibitem{hill1966effect}
Hill WG, Robertson A.
\newblock The effect of linkage on limits to artificial selection.
\newblock Genetical research. 1966;8(03):269--294.

\bibitem{eichler2003structural}
Eichler EE, Sankoff D.
\newblock Structural dynamics of eukaryotic chromosome evolution.
\newblock Science. 2003;301(5634):793--797.

\bibitem{comeron2008hill}
Comeron JM, Williford A, Kliman R.
\newblock The Hill--Robertson effect: evolutionary consequences of weak
  selection and linkage in finite populations.
\newblock Heredity. 2008;100(1):19--31.

\bibitem{presgraves2005recombination}
Presgraves DC.
\newblock Recombination enhances protein adaptation in Drosophila melanogaster.
\newblock Current Biology. 2005;15(18):1651--1656.

\bibitem{barton2010genetic}
Barton N.
\newblock Genetic linkage and natural selection.
\newblock Philosophical Transactions of the Royal Society of London B:
  Biological Sciences. 2010;365(1552):2559--2569.

\bibitem{magnusson2012demasculinization}
Magnusson K, Lycett GJ, Mendes AM, Lynd A, Papathanos PA, Crisanti A, et~al.
\newblock Demasculinization of the Anopheles gambiae X chromosome.
\newblock BMC evolutionary biology. 2012;12(1):69.

\bibitem{jaquiery2013masculinization}
Jaqui{\'e}ry J, Rispe C, Roze D, Legeai F, Le~Trionnaire G, Stoeckel S, et~al.
\newblock Masculinization of the X chromosome in the pea aphid.
\newblock Plos Genet. 2013;9(8):e1003690.

\bibitem{quail2012tale}
Quail MA, Smith M, Coupland P, Otto TD, Harris SR, Connor TR, et~al.
\newblock A tale of three next generation sequencing platforms: comparison of
  Ion Torrent, Pacific Biosciences and Illumina MiSeq sequencers.
\newblock BMC genomics. 2012;13(1):341.

\bibitem{dohm2008substantial}
Dohm JC, Lottaz C, Borodina T, Himmelbauer H.
\newblock Substantial biases in ultra-short read data sets from high-throughput
  DNA sequencing.
\newblock Nucleic acids research. 2008;36(16):e105--e105.

\bibitem{balzer2013filtering}
Balzer S, Malde K, Grohme MA, Jonassen I.
\newblock Filtering duplicate reads from 454 pyrosequencing data.
\newblock Bioinformatics. 2013;29(7):830--836.

\bibitem{hansen2010biases}
Hansen KD, Brenner SE, Dudoit S.
\newblock Biases in Illumina transcriptome sequencing caused by random hexamer
  priming.
\newblock Nucleic acids research. 2010;38(12):e131--e131.

\bibitem{brown2012extensive}
Brown KH, Dobrinski KP, Lee AS, Gokcumen O, Mills RE, Shi X, et~al.
\newblock Extensive genetic diversity and substructuring among zebrafish
  strains revealed through copy number variant analysis.
\newblock Proceedings of the National Academy of Sciences.
  2012;109(2):529--534.

\bibitem{li2009fast}
Li H, Durbin R.
\newblock Fast and accurate short read alignment with Burrows--Wheeler
  transform.
\newblock Bioinformatics. 2009;25(14):1754--1760.

\bibitem{langmead2012fast}
Langmead B, Salzberg SL.
\newblock Fast gapped-read alignment with Bowtie 2.
\newblock Nature methods. 2012;9(4):357--359.

\bibitem{simpson2012efficient}
Simpson JT, Durbin R.
\newblock Efficient de novo assembly of large genomes using compressed data
  structures.
\newblock Genome research. 2012;22(3):549--556.

\bibitem{liu2013musket}
Liu Y, Schr{\"o}der J, Schmidt B.
\newblock Musket: a multistage k-mer spectrum-based error corrector for
  Illumina sequence data.
\newblock Bioinformatics. 2013;29(3):308--315.

\bibitem{chen2014tigra}
Chen K, Chen L, Fan X, Wallis J, Ding L, Weinstock G.
\newblock TIGRA: a targeted iterative graph routing assembler for breakpoint
  assembly.
\newblock Genome research. 2014;24(2):310--317.

\bibitem{malde2006rbr}
Malde K, Schneeberger K, Coward E, Jonassen I.
\newblock RBR: library-less repeat detection for ESTs.
\newblock Bioinformatics. 2006;22(18):2232--2236.

\bibitem{brown2012reference}
Brown CT, Howe A, Zhang Q, Pyrkosz AB, Brom TH.
\newblock A reference-free algorithm for computational normalization of shotgun
  sequencing data.
\newblock arXiv preprint arXiv:12034802. 2012;.

\bibitem{malde2014estimating}
Malde K.
\newblock Estimating the information value of polymorphic sites using pooled
  sequences.
\newblock BMC genomics. 2014;15(Suppl 6):S20.

\bibitem{chen2012identification}
Chen N, Bellott DW, Page DC, Clark AG.
\newblock Identification of avian W-linked contigs by short-read sequencing.
\newblock BMC genomics. 2012;13(1):183.

\bibitem{hall2013six}
Hall AB, Qi Y, Timoshevskiy V, Sharakhova MV, Sharakhov IV, Tu Z.
\newblock Six novel Y chromosome genes in Anopheles mosquitoes discovered by
  independently sequencing males and females.
\newblock Bmc Genomics. 2013;14(1):273.

\bibitem{carvalho2013efficient}
Carvalho AB, Clark AG.
\newblock Efficient identification of Y chromosome sequences in the human and
  Drosophila genomes.
\newblock Genome research. 2013;23(11):1894--1907.

\bibitem{tomaszkiewicz2017and}
Tomaszkiewicz M, Medvedev P, Makova KD.
\newblock Y and W chromosome assemblies: approaches and discoveries.
\newblock Trends in Genetics. 2017;.

\bibitem{fraisse2017deep}
Fra{\"\i}sse C, Picard MA, Vicoso B.
\newblock The deep conservation of the Lepidoptera Z chromosome suggests a
  non-canonical origin of the W.
\newblock Nature communications. 2017;8(1):1486.

\bibitem{vicoso2013comparative}
Vicoso B, Emerson J, Zektser Y, Mahajan S, Bachtrog D.
\newblock Comparative sex chromosome genomics in snakes: differentiation,
  evolutionary strata, and lack of global dosage compensation.
\newblock PLoS Biology. 2013;11(8):e1001643.

\bibitem{vicoso2013reversal}
Vicoso B, Bachtrog D.
\newblock Reversal of an ancient sex chromosome to an autosome in Drosophila.
\newblock Nature. 2013;499(7458):332.

\bibitem{kishida2015aquatic}
Kishida T, Thewissen J, Hayakawa T, Imai H, Agata K.
\newblock Aquatic adaptation and the evolution of smell and taste in whales.
\newblock Zoological Letters. 2015;1(1):1--10.

\bibitem{malde2017whole}
Malde K, Seliussen BB, Quintela M, Dahle G, Besnier F, Skaug HJ, et~al.
\newblock Whole genome resequencing reveals diagnostic markers for
  investigating global migration and hybridization between minke whale species.
\newblock BMC genomics. 2017;18(1):76.

\bibitem{li2009sequence}
Li H, Handsaker B, Wysoker A, Fennell T, Ruan J, Homer N, et~al.
\newblock The sequence alignment/map format and SAMtools.
\newblock Bioinformatics. 2009;25(16):2078--2079.

\bibitem{berube1996identification}
B{\'e}rub{\'e} M, Palsboll P.
\newblock Identification of sex in cetaceans by multiplexing with three ZFX and
  ZFY specific primers.
\newblock Molecular Ecology. 1996;5(2):283--287.

\bibitem{harris2007improved}
Harris RS.
\newblock Improved pairwise alignment of genomic DNA.
\newblock ProQuest; 2007.

\bibitem{ross2005dna}
Ross MT, Grafham DV, Coffey AJ, Scherer S, McLay K, Muzny D, et~al.
\newblock The DNA sequence of the human X chromosome.
\newblock Nature. 2005;434(7031):325--337.

\bibitem{pastene2007radiation}
Pastene LA, Goto M, Kanda N, Zerbini AN, Kerem D, Watanabe K, et~al.
\newblock Radiation and speciation of pelagic organisms during periods of
  global warming: the case of the common minke whale, Balaenoptera
  acutorostrata.
\newblock Molecular Ecology. 2007;16(7):1481--1495.

\bibitem{arnason1993cetacean}
Arnason U, Gullberg A, Widegren B.
\newblock Cetacean mitochondrial DNA control region: sequences of all extant
  baleen whales and two sperm whale species.
\newblock Molecular Biology and Evolution. 1993;10(5):960--970.

\bibitem{gamble2015restriction}
Gamble T, Coryell J, Ezaz T, Lynch J, Scantlebury DP, Zarkower D.
\newblock Restriction site-associated DNA sequencing (RAD-seq) reveals an
  extraordinary number of transitions among gecko sex-determining systems.
\newblock Molecular Biology and Evolution. 2015;32(5):1296--1309.

\bibitem{rovatsos2016mammalian}
Rovatsos M, Vuki{\'c} J, Kratochv{\'\i}l L.
\newblock Mammalian X homolog acts as sex chromosome in lacertid lizards.
\newblock Heredity. 2016;117(1):8.

\bibitem{nla.cat-vn1871086}
White MJD.
\newblock The chromosomes [by] M. J. D. White.
\newblock 6th ed. Chapman and Hall, distributed by Halsted Press, New York
  London; 1973.

\end{thebibliography}

\end{document}